\documentclass[final]{svjour3}
\usepackage{graphicx}
\usepackage{rotating}
\usepackage{amssymb}
\usepackage{mathptmx}
\usepackage[numbers]{natbib}
\usepackage{xcolor}
\usepackage{comment}
\makeatletter
\journalname{Journal of Low Temperature Physics}

\begin{document}

\title{Characterization of Transition Edge Sensors for the Simons Observatory}

\author{Jason R. Stevens\textsuperscript{1}\kern-1.5pt \and 
Nicholas F. Cothard\textsuperscript{1}\kern-1.5pt \and 
Eve M. Vavagiakis\textsuperscript{1}\kern-1.5pt \and
Aamir Ali\textsuperscript{2}\kern-1.5pt \and
Kam Arnold\textsuperscript{3}\kern-1.5pt \and
Jason E. Austermann\textsuperscript{4}\kern-1.5pt \and 
Steve K. Choi\textsuperscript{1}\kern-1.5pt \and 
Bradley J. Dober\textsuperscript{4}\kern-1.5pt \and 
Cody Duell\textsuperscript{1}\kern-1.5pt \and
Shannon M. Duff\textsuperscript{4}\kern-1.5pt \and 
Gene C. Hilton\textsuperscript{4}\kern-1.5pt \and 
Shuay-Pwu Patty Ho\textsuperscript{5}\kern-1.5pt \and
Thuong D. Hoang\textsuperscript{1}\kern-1.5pt \and 
Johannes Hubmayr\textsuperscript{4}\kern-1.5pt \and 
Adrian T. Lee\textsuperscript{2}\kern-1.5pt \and 
Aashrita Mangu\textsuperscript{2}\kern-1.5pt \and
Federico Nati\textsuperscript{6}\kern-1.5pt \and
Michael D. Niemack\textsuperscript{1}\kern-1.5pt \and 
Christopher Raum\textsuperscript{2}\kern-1.5pt \and 
Mario Renzullo\textsuperscript{7}\kern-1.5pt \and 
Maria Salatino\textsuperscript{8}\kern-1.5pt \and
Trevor Sasse\textsuperscript{2}\kern-1.5pt \and 
Sara M. Simon\textsuperscript{9} \and
Suzanne Staggs\textsuperscript{5} \and
Aritoki Suzuki\textsuperscript{2,7}\kern-1.5pt \and 
Patrick Truitt\textsuperscript{7}\kern-1.5pt \and 
Joel Ullom\textsuperscript{4}\kern-1.5pt \and 
John Vivalda\textsuperscript{7}\kern-1.5pt \and 
Michael R. Vissers\textsuperscript{4}\kern-1.5pt \and 
Samantha Walker\textsuperscript{4}\kern-1.5pt \and 
Benjamin Westbrook\textsuperscript{2}\kern-1.5pt \and 
Edward J. Wollack\textsuperscript{10}\kern-1.5pt \and
Zhilei Xu\textsuperscript{11}\kern-1.5pt \and
Daniel Yohannes\textsuperscript{7}\kern-1.5pt
}

\institute{\footnotesize
\noindent\textsuperscript{1}Cornell University, Ithaca, NY, USA \\
\noindent\textsuperscript{2}University of California Berkeley, Berkeley, CA, USA \\
\noindent\textsuperscript{3}University of California San Diego, San Diego, CA, USA \\
\noindent\textsuperscript{4}NIST, Boulder, CO, USA \\
\noindent\textsuperscript{5}Princeton University, Princeton, NJ, USA \\
\noindent\textsuperscript{6}University of Milano - Bicocca, Italy \\
\noindent\textsuperscript{7}HYPRES/SeeQC, Elmsford, NY, USA \\
\noindent\textsuperscript{8}KIPAC/Stanford University, Standford, CA, USA \\
\noindent\textsuperscript{9}University of Michigan, Ann Arbor, MI, USA \\
\noindent\textsuperscript{10}NASA/Goddard Space Flight Center, Greenbelt, MD, USA \\
\noindent\textsuperscript{11}University of Pennsylvania, Philadelphia, PA, USA
}



\maketitle

\begin{abstract}
The Simons Observatory is building both large (6 m) and small (0.5 m) aperture telescopes in the Atacama desert in Chile to observe the cosmic microwave background (CMB) radiation with unprecedented sensitivity. Simons Observatory telescopes in total will use over 60,000 transition edge sensor (TES) detectors spanning center frequencies between 27 and 285 GHz and operating near 100 mK.                                                                                         
TES devices have been fabricated for the Simons Observatory by NIST, Berkeley, and HYPRES/SeeQC corporation. Iterations of these devices have been tested cryogenically in order to inform the fabrication of further devices, which will culminate in the final TES designs to be deployed in the field. The detailed design specifications have been independently iterated at each fabrication facility for particular detector frequencies.                               

We present test results for prototype devices, with emphasis on NIST high frequency detectors. A dilution refrigerator was used to achieve the required temperatures. Measurements were made both with 4-lead resistance measurements and with a time domain Superconducting Quantum Interference Device (SQUID) multiplexer system. The SQUID readout measurements include analysis of current vs voltage (IV) curves at various temperatures, square wave bias step measurements, and detector noise measurements. Normal resistance, superconducting critical temperature, saturation power, thermal and natural time constants, and thermal properties of the devices are extracted from these measurements.
\end{abstract}

\section{Introduction}

The Simons Observatory collaboration is building a series of telescopes to measure the CMB radiation; one large aperture ($\sim$6m) telescope and three small aperture ($\sim$0.5 m) telescopes \cite{galitz}.
Each Simons Observatory telescope will use arrays of TES bolometers to measure power from the CMB at multiple frequencies. In total, over 60,000 such sensors will be used in the polarization sensitive focal plane \cite{orlow}. The Simons Observatory target bands are ``low frequency'' LF-1 $\sim$27 GHz and LF-2 $\sim$39 GHz, ``Medium Frequency'' MF-1 $\sim$93 GHz and MF-2 $\sim$145 GHz, and ``ultra high frequency'' UHF-1 $\sim$225 GHz and UHF-2 $\sim$285 GHz.

A TES consists of a superconducting material thermally linked to a constant temperature bath. The TES is voltage biased to keep it on its superconducting transition, where the resistance of the device changes steeply with input power. Voltage biasing the TES results in negative electro-thermal feedback that keeps the TES stably on the superconducting transition \cite{irwinhilton}. A resistance change due to power incident on the detector from the sky changes the current through the detector; this current can be measured with any one of a number of SQUID based readout systems \cite{irwinhilton}\cite{mates}.

A TES has a number of properties that can be optimized according to the desired application. The observation frequency, temperature of the thermal bath, the SQUID readout architecture, and expected power from the atmosphere all affect target TES parameters. Therefore, iterative testing is underway to develop fabrication processes for the TES bolometers that will be used in the Simons Observatory telescopes.

Several TES parameters must be optimized to match the planned readout and cryogenic systems. We target a normal resistance ($R_n$) of 8 m$\Omega$ to match the microwave SQUID multiplexing readout system described in \cite{henderson}. Lowering the resistance increases the power-to-current responsivity of the TES, which results in the SQUID readout noise contribution being suppressed relative to the TES noise. Stability is maintained by operating the TESes in the overdamped regime in which the electrical time constants of the circuit are much shorter than the TES thermal time constants \cite{irwinhilton}. For these tests, we operate with a shunt resistance of 200$\mu \Omega$. 
We target a critical temperature ($T_c$) of 160 mK for use with dilution-refrigerator-cooled cameras. Other significant band dependent target parameters are listed in Table \ref{tab:targets}.

\begin{table}
    \centering
    \begin{tabular}{| l | c | c | r |}
        \hline
        Frequency & Parameter & Target & Variation \\ \hline
        MF-1 (93 GHz)    & $P_{sat}$(100mK)    & 4 pW                  & 3-5 pW \\
                         & $\tau_{eff}$        & 0.61ms                & 0.37-1.1 ms \\
        MF-2 (145 Ghz)   & $P_{sat}$(100mK)    & 6.3 pW                & 4.7 - 7.9 pW \\
                         & $\tau_{eff}$         & 0.53 ms               & 0.32-0.96 ms \\
        UHF-1  (225 GHz) & $P_{sat}$(100mK)   & 16 pW              & 12-19 pW \\
                         & $\tau_{eff}$       & 0.36ms               & 0.2-0.65 ms \\
        UHF-2 (285 GHz)  & $P_{sat}$(100mK)   & 24 pW              & 18-31 pW \\
                         & $\tau_{eff}$       & 0.31ms               & 0.18-0.57 ms \\
        \hline
    \end{tabular}
    \caption{Current SO targets for some of the measured parameters in various bands. $P_{sat}$ targets, are chosen based on loading estimation for bands; noise targets are motivated by sensitivity requirements; $\tau_{eff}$ targets are motivated by the expected rate of change of the TES input signal. We have omitted the low frequency (LF) bands as their target parameters have not been sufficiently determined as of this writing.}
    \label{tab:targets}
\end{table}

To inform each new generation of TES fabriaction for the Simons Observatory, TES properties were tested at Cornell University. All measurements were done in a dilution refrigerator with a mixing chamber heater capable of servoing the temperature to a specified value. All measurements were done in a dark environment; that is, no TES was exposed to significant external light. Except for the four lead measurements, all measurements were taken with the aid of a time domain SQUID multiplexing system, read out through Multi-Channel Electronics (MCE) developed at the University of British Columbia \cite{battistelli}. The MCE allows us to specify the TES bias voltage and read out the resulting current through the TES. The time domain multiplexing system employed with the MCE is a good proxy for the microwave multiplexing that will be used in Simons Observatory because the TES bias circuit is the same and the measurements are not readout noise dominated (see Sec. \ref{sec:noise}, Fig. \ref{fig:noise}, and Fig. \ref{fig:nep_hist}).

We present characterization of detectors fabricated at NIST, Berkeley, and SeeQC. The SeeQC detectors are also described in \cite{toki}. 
Due to fabrication heritage, the detectors fabricated at Berkeley and SeeQC have the same design as each other, but do not share design elements with the NIST detectors. This is in part due to the different approaches used to thermally isolate the TESes -- deep reactive ion etching is used at NIST, while Xenon diflouride etching is used at Berkeley and SeeQC. Therefore, measurements of NIST detectors do not directly inform the Berkeley/SeeQC design or fabrication, and vice versa. Given that this is the first time the teams at Berkeley and SeeQC have fabricated AlMn TESes with target resistance of $R_n= 0.008$\,$\Omega$ and $T_c = 0.16$\,K, results from these devices are less mature. Therefore, only four lead and IV measurements of them are presented at this time.
Devices from all facilities were fabricated with AlMn from ACI Alloys, Inc. The required annealing temperature to achieve the target Tc \cite{li} varies between fabrication facilities, which is not surprising given the different TES geometries and fabrication process differences (e.g. DRIE vs. XeF2 etching) between the facilities.

The NIST detectors presented here have evolved from the designs optimized for Advanced ACTPol \cite{choi} and are currently being optimized for UHF band observations. Two generations of these detectors have been fabricated and tested, and are hereafter referred to as v1 and v2. Feedback from the testing presented here has informed the optimization of the v2 detectors, which we show are closer to the target SO parameters. However, additional iterations are planned for all devices to improve the performance further.

\section{Four lead and IV Measurements}


\begin{figure}
    \centering
    \includegraphics[height=1.15in]{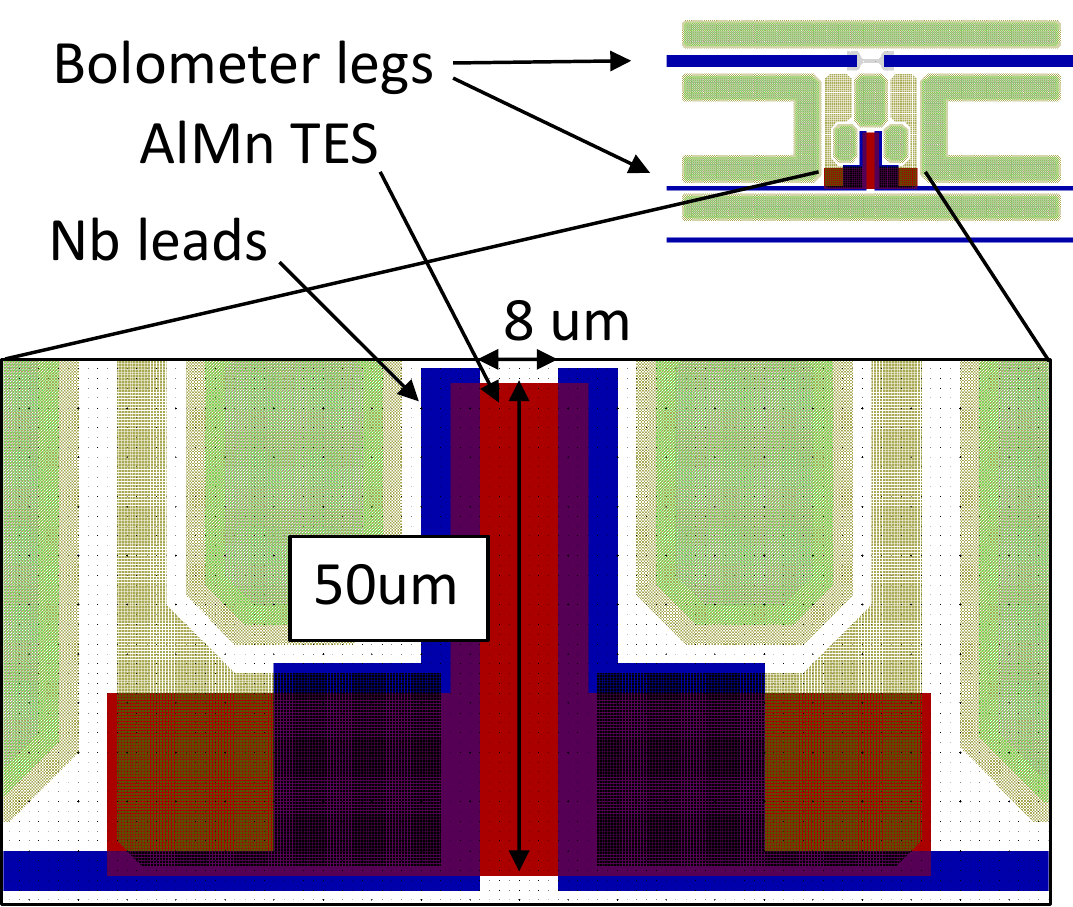} 
    \includegraphics[height=1.15in]{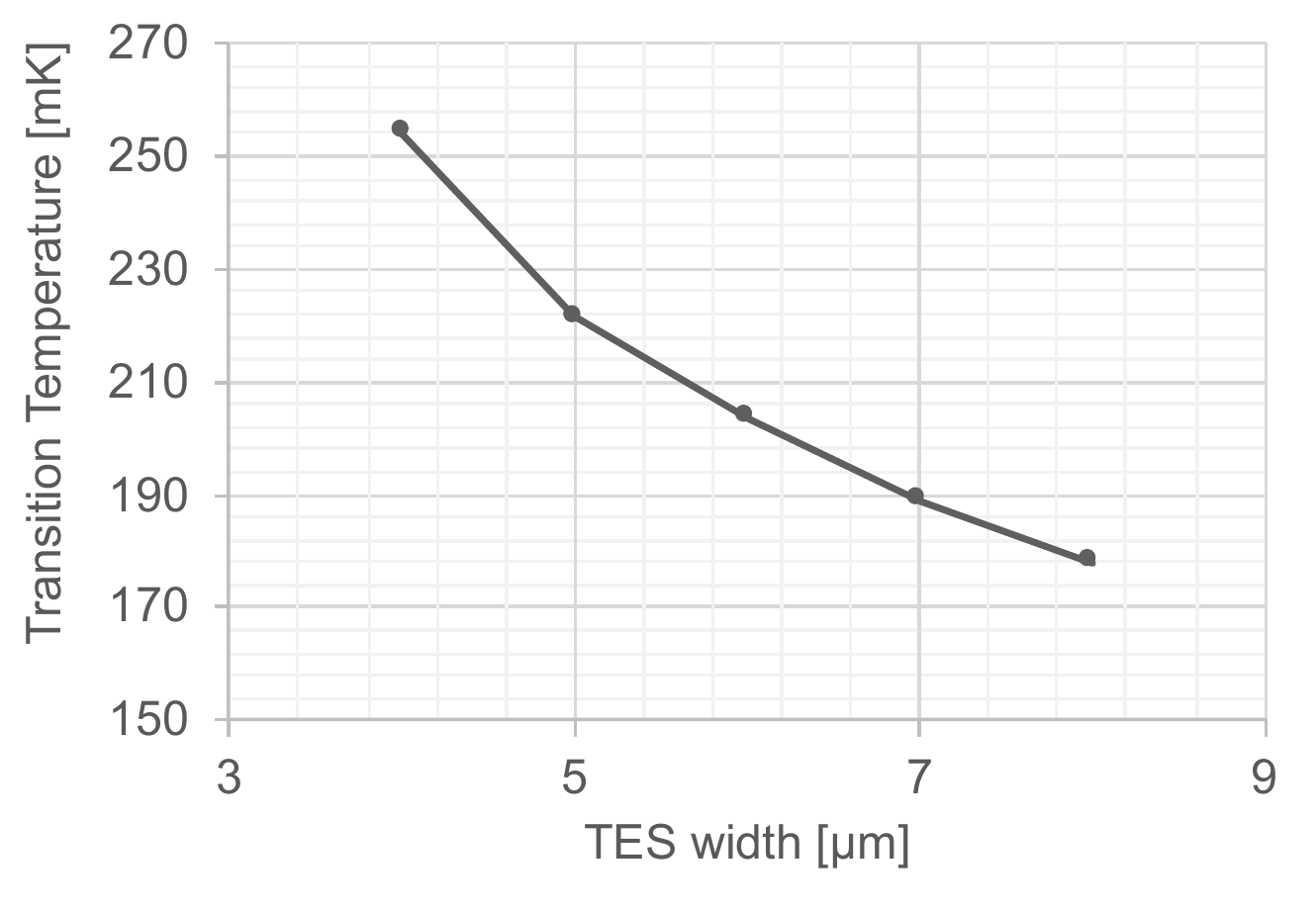} 
    \includegraphics[height=1.15in]{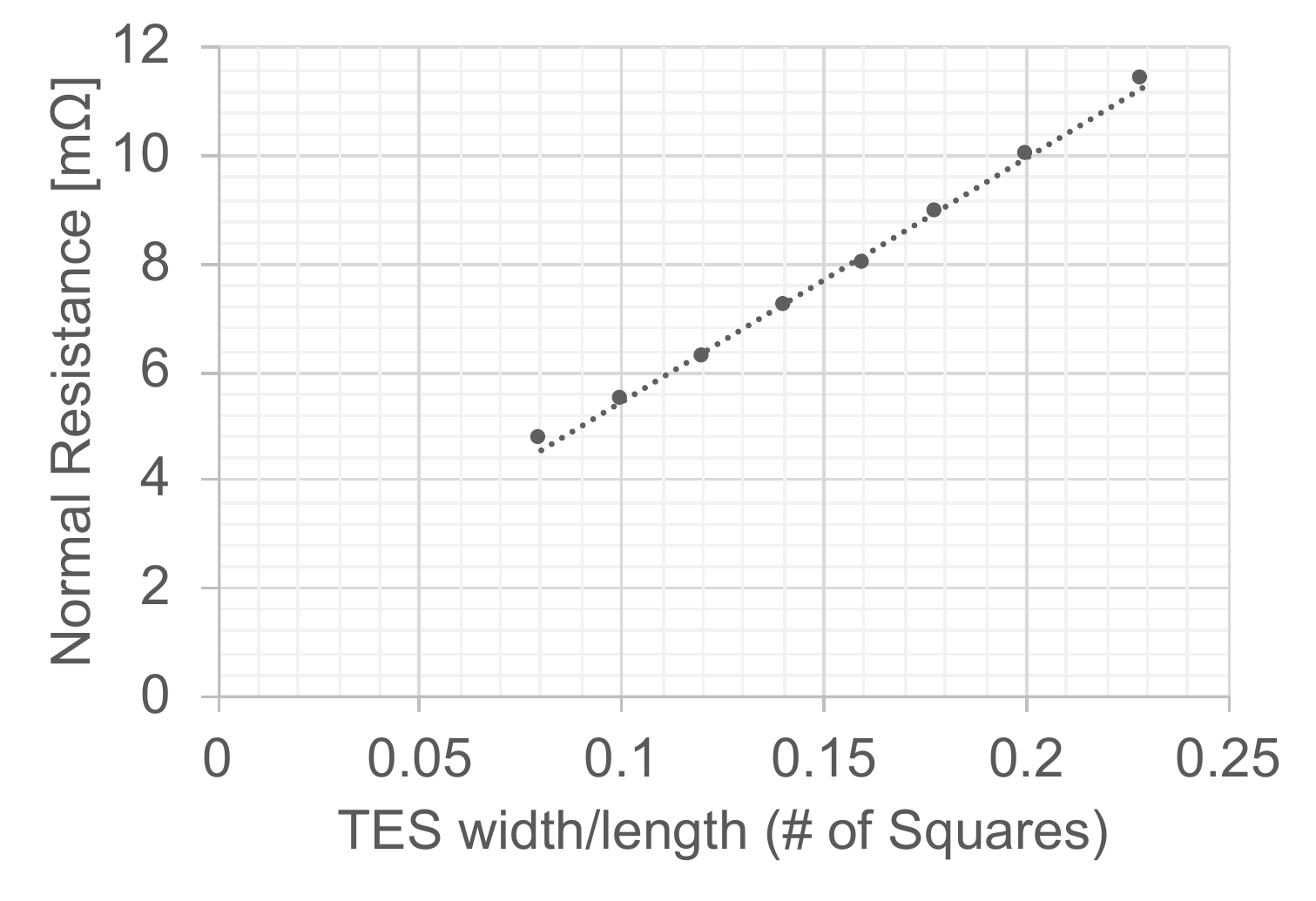} 
 
    
    \caption{{\it Left}: SeeQC bolometer and TES design highlighting the bolometer legs that are adjusted to change the saturation power (Fig.~\ref{fig:ivana}), the TES geometry (red), which is 8\,$\mu$m by 50\,$\mu$m in this design, and the Nb leads. {\it Middle}: TES $T_c$ versus width. The increase in $T_c$ for smaller widths is attributed to an increase in the proximity effect from the Nb leads. {\it Right}: TES $R_n$ versus the ratio of the TES width to the length (aka. number of squares in the TES).}
    \label{fig:hypres}
\end{figure}


Using the MCE, the current through the TES can be measured as a function of bias voltage across the superconducting transition. This allows measurement of the saturation power required to drive the TES normal. By measuring the saturation power at many different temperatures, we can fit the model

\begin{equation}
P_{sat} = k ( T_c^n - T_{bath}^n ).
\end{equation}

\begin{figure}
    \centering
\begin{tabular}{l r}
    \includegraphics[scale=0.35]{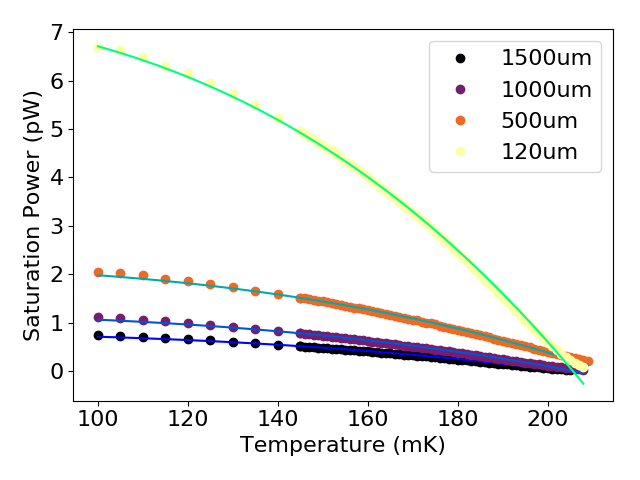} & \includegraphics[scale=0.35]{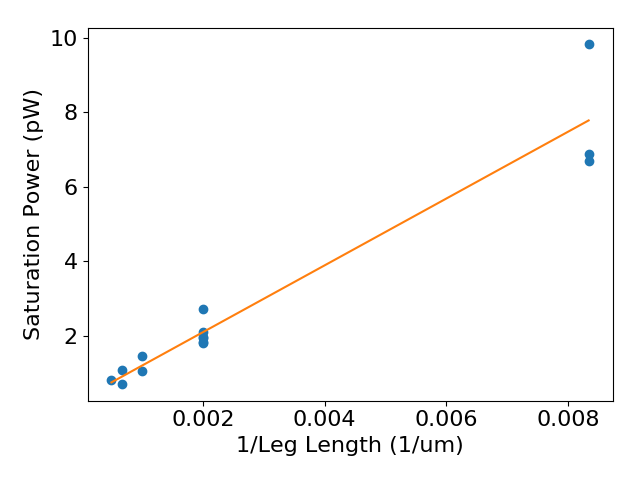} \\
\end{tabular}
    
    \caption{{\it Left}: Example fit to $P_{sat}$ vs bath temperature curve for a set of Berkeley detectors with various leg lengths. Each point represents saturation power extracted from an I-V curve at a fixed temperature, and the model allows extraction of critical temperature and thermal conductance. {\it Right}: Saturation power at 100 mK vs inverse leg length for a number of Berkeley detectors with linear regression. The saturation power should go as cross sectional area over leg length \cite{koopman}; in these detectors, the cross sectional area is constant.}
    \label{fig:ivana}
\end{figure}

By doing so we measure both $T_c$ and $G = nkT^{n-1}$, which is the thermal conductance between the TES and the bath \cite{irwinhilton} (Fig. \ref{fig:ivana}). Additionally, the saturation power itself is a significant target parameter, as it must be tuned to match the expected optical loading from the sky at each frequency. Analysis of the IVs collected as part of this process also naturally yields measurements of $R_n$.

The variety of information yielded by this analysis makes it a critical component of TES testing. Using this method, we successfully measured the parameters of the NIST UHF detectors shown in Tab. \ref{tab:nist_uhf_results}. The feedback from these tests on the v1 detectors led to the improvement seen in the v2 detectors.

\begin{table}
    \centering
    \begin{tabular}{|l|c|c|c|}
    \hline
         Parameter & Target & Measured, v1 & Measured, v2 \\
         \hline
         $T_c$              & 160 mK        & 186 mK & 166 mK \\
         $P_{sat}$ 225 GHz  & 12-19 pW  & 26 pW & 18 pW \\
         $P_{sat}$ 285 GHz  & 18-31 pW  & 30 pW & 24 pW \\
         $R_n$ 225 GHz      & 8 m$\Omega$     & 7.1 m$\Omega$ & 7.8 m$\Omega$ \\
         $R_n$ 285 GHz      & 8 m$\Omega$     & 7.6 m$\Omega$ & 7.9 m$\Omega$ \\
    \hline
    \end{tabular}
    \caption{Some measured parameters from NIST UHF detectors compared to their target ranges as gathered from IV analysis. $P_{sat}$ values are listed for a bath temperature of 100mK. Measurements are presented for both v1 and v2 detectors. In each case, the v2 value is closer to the target.}
    \label{tab:nist_uhf_results}
\end{table}

SQUID-based measurements of $T_c$ and $R_n$ were confirmed with extensive four lead resistance measurements. Four lead measurements are faster and easier to acquire, and can reveal an appropriate range of temperatures to take IV measurements over for a given device. Example measurements of several SeeQC detectors are shown in Fig.~\ref{fig:hypres} for different TES geometries. More information about these devices is available in~\cite{toki}. Variations in $T_c$ are attributed to changes in the proximity effect from the Nb leads. $R_n$ scales with geometry as expected (Fig.~\ref{fig:hypres}). Similar behaviors have been observed with previous AlMn TESes \cite{li}.

We also measured a variety of TES devices fabricated by Berkeley in order to inform leg geometry for their specific detector design (Fig. \ref{fig:ivana}). Berkeley recently installed a new AlMn target and is continuing to optimize $T_c$ and $R_n$. This optimization combined with the information in Fig. \ref{fig:ivana} will be used to select TES leg geometries that achieve the desired saturation power for the next round of Berkeley devices.

\section{Bias Step Measurements}
\label{ref:biasstepsection}

\begin{figure}
    \centering
\begin{tabular}{l r}
    \includegraphics[width=0.47\linewidth]{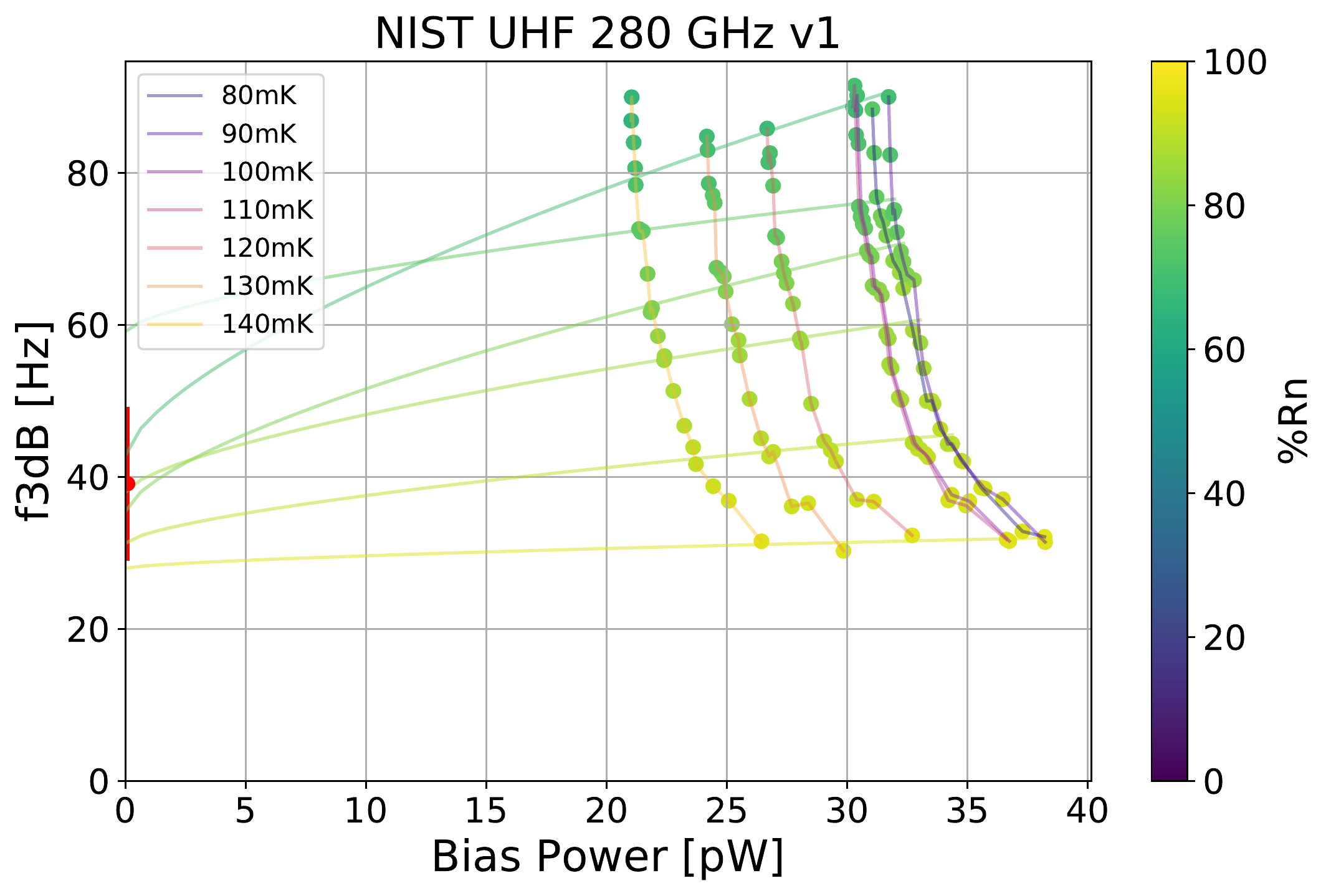} & \includegraphics[width=0.47\linewidth]{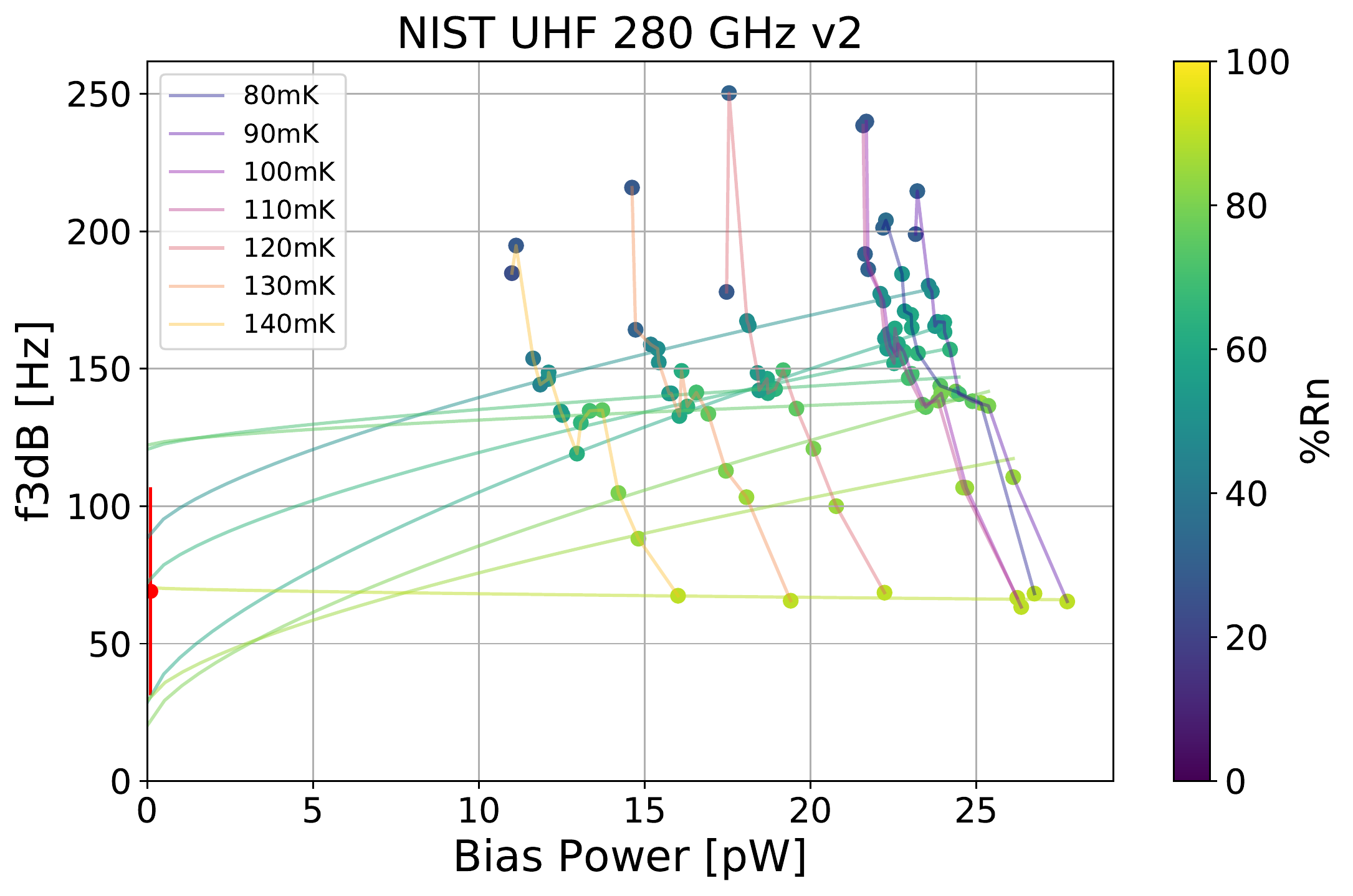} \\
\end{tabular}
    \caption{Measured values of $f_{3dB} = 1/2 \pi \tau_{eff}$ for NIST UHF-2 (285 GHz) detectors. These data were taken with the methods described in Section \ref{ref:biasstepsection} at various temperatures and fractions of normal resistance. For a set of measurements with the same fraction of normal resistance, a fit to Equation \ref{eq:two-fuid} is used to extrapolate to zero bias power, in order to estimate the natural time constant $f_{nat}$. {\it Left}: Measurements of a v1 detector are slower than the target design, which motivated the removal of heat capacity on the bolometer island. {\it Right}: Measurements of a v2 detector which showed a faster response, as expected after heat capacity removal, although still slower than the target design.}
    \label{fig:f3db}
\end{figure}

The TES temporal response while operated under negative electrothermal feedback can be quantified by  $f_{3dB}$, which is related to the detector's effective thermal time constant $f_{3dB} = 1 / 2 \pi \tau_{eff}$. We measure $f_{3dB}$ by first biasing the TES onto its transition and then applying a small amplitude square wave on top of the DC bias. The TES response to the square wave is sampled quickly ($\sim3200\textrm{ Hz}$) and the time constant is extracted via a single pole exponential fit to each step of the square wave \cite{koopman}.

Fig. \ref{fig:f3db} shows measurements of the thermal time constants of NIST UHF detectors. These measurements have been performed at multiple bath temperatures and multiple points on the transition (shown as fraction of normal resistance). These data are then fit (for a given fraction of normal resistance) to
\begin{equation}
\label{eq:two-fuid}
f_{3dB} = A + B P_{bias}^{{2} \over {3}}
\end{equation}
where $A$ and $B$ are a function of the measurable parameters \cite{two-fluid}. The natural time constant, $\tau_{nat} = C/G$, is equivalent to the bolometer time constant without negative electrothermal feedback and is extrapolated from the value of $f_{3dB}$ at zero power:  $f_{nat} = (1/2\pi)G/C$. 

In Fig. \ref{fig:f3db}, we show bias step $f_{3dB}$ measurements of v1 and v2 devices. Fits to Eq. \ref{eq:two-fuid} are extrapolated to zero bias power and the average $f_{nat}$ of the fits is plotted with an errorbar corresponding to the standard deviation between fits. In the operational bias power range, the time constants of v1 devices were found to be too slow, motivating the removal of some of the bolometer heat capacity in v2 devices. The bolometer heat capacity is dominated ($\sim99\%$) by the TES AlMn and extra PdAu used to control heat capacity volume. Table \ref{tab:f3dB} gives the fabricated volumes of the PdAu and AlMn heat capacities. Given these volumes and that the heat capacity per unit volume of PdAu is roughly 3.5 greater than that of AlMn \cite{C_PdAu} combined with measurements of $\rm T_c$ and G, we calculate that $f_{nat}$ for v2 should be roughly twice that of v1 (Table~\ref{tab:f3dB}). While this prediction does not account for the loop gain of the TES, and thus does not indicate the speed-up of the devices under negative electrothermal feedback, it provides a guide and estimation for the next iteration of fabricated devices. Indeed, after measurements, the extrapolated $f_{nat}$ of v2 was found to be roughly twice that of v1, as shown in Fig. \ref{fig:f3db}. Future v3 TESes are being fabricated with $\sim52\%$ less PdAu volume than v2. This is expected to decrease the natural time constants by an additional factor of 1.5 between the v2 and v3 detectors.


\begin{table}
    \centering
    \begin{tabular}{|r|c|c|c|c|}
    \hline
         Detector    & $\rm V_{PdAu,Tc160mK}$ & $ \rm V_{AlMn,Tc160mK}$ & $\rm \tau_{nat} \propto V_{eff}/G$ & $\tau_{nat,v1}/\tau_{nat,v2}$ \\
         \hline
         285 GHz v1  & $\rm 45,000 ~\mu m^3$ & $\rm 37,000 ~\mu m^3$ & $\rm 382,000 ~\mu m^3 W K^{-1}$ &  \\
         285 GHz v2  & $\rm 18,800 ~\mu m^3$ & $\rm 37,000 ~\mu m^3$ & $\rm 186,000 ~\mu m^3 W K^{-1}$ & 2.1 \\
    \hline
    \end{tabular}
    \caption{Fabricated PdAu and AlMn volumes used to estimate the speed-up between v1 and v2 time constants. The speed-up factor is calculated by taking the ratio of $\tau_{nat}=C/G$ of v1 and v2. An effective heat capacity is calculated by given that the heat capacity per unit volume of PdAu is roughly 3.5 greater than AlMn, and assuming that the heat capacity scales linearly with critical temperature: $V_{eff}=(3.5 \cdot V_{PdAu}+V_{AlMn}) \cdot (T_c/160\rm mK)$. Then, $V_{eff}/G$ is used as a proxy for the $\tau_{nat}$ and the speed-up factor is taken as the ratio of the proxies for v1 and v2.}
    
    \label{tab:f3dB}
\end{table}

\section{Noise Measurements}
\label{sec:noise}
The readout system is designed to bias the TES onto the superconducting transition and then read out the current as a function of time, while keeping the TES on the transition. In the field, this current will be used to measure input power due to light from the sky. However, dark measurements in the laboratory are useful for measuring the noise characteristics of the TES.

In our tests, data are sampled at 3200 Hz for one minute. These data streams are acquired on each detector at many different temperatures and points on the superconducting transition (i.e., fractions of normal resistance).

\begin{figure}
    \centering
    \begin{tabular}{c c}
         &  \\
         & 
    \end{tabular}
    \includegraphics[scale=0.4]{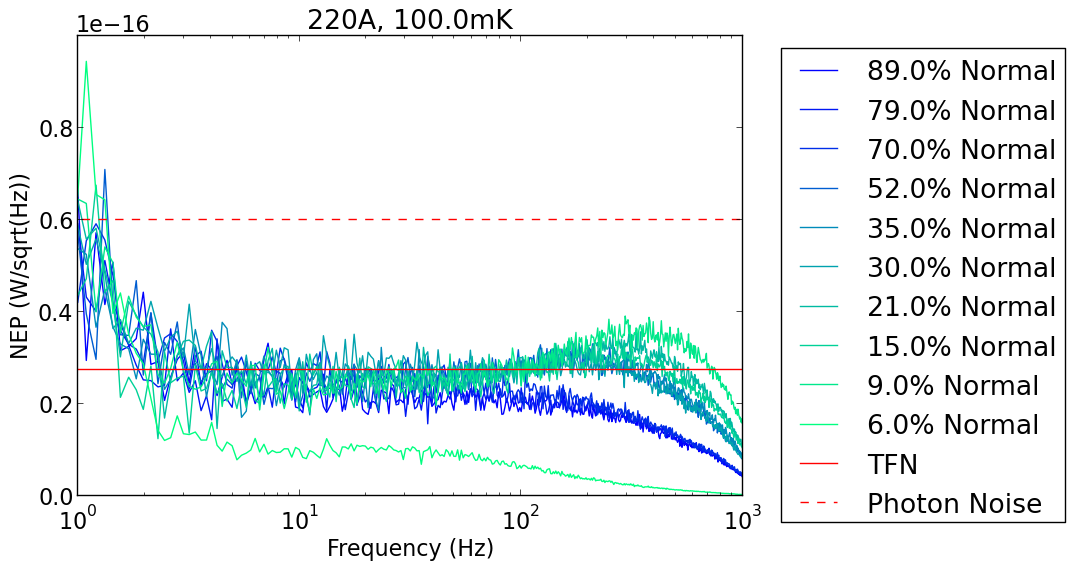}
    \caption{Noise measurements for one NIST UHF-1 v2 TES at various percentages of normal resistance, at 100\,mK.  The data were sampled at 3200\,Hz. A DC approximation of thermal fluctuation noise (TFN) is shown (solid red line) and is consistent with the measured noise level between a few to 100\,Hz. The photon noise level for UHF-1 on the SAT is expected to be greater than 60\,aW/Hz$^{1/2}$ (dashed red line) \cite{so_forecast}. 
    The fact that the measured dark detector noise is significantly less than the expected photon noise suggests that these detectors will be photon-noise limited when deployed.}
    \label{fig:noise}
\end{figure}

\begin{figure}
    \includegraphics[scale=0.5]{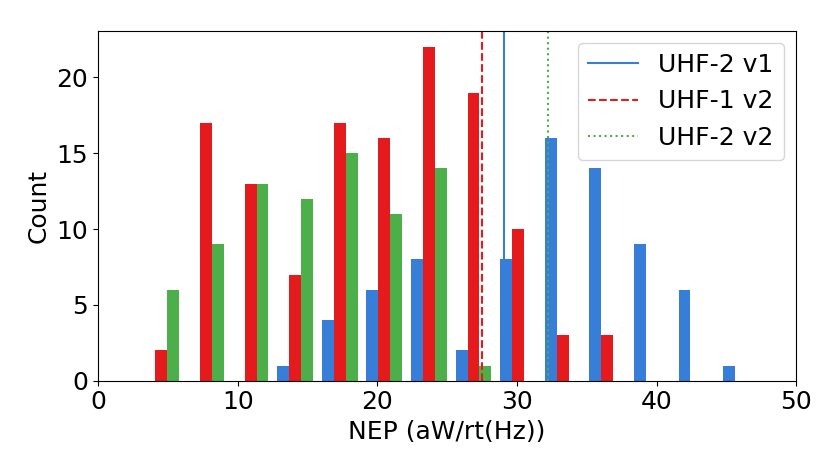}
    \caption{Histograms of measured NEP for NIST UHF detectors. In these histograms, each detector was measured at various temperatures and fractions of normal resistance, and each measurement counts as a point on the histogram. In total, there are six independent physical detectors measured. The thermal fluctuation noise (TFN) is estimated and plotted as a vertical line for each detector type. The measured noise levels cluster around (v1) or slightly below (v2) the TFN, suggesting it is the dominant noise source. The TFN is calculated by using the average measured $G$ and $T_c$ of the devices.  The NEP of the detector is a fit of a constant function to the NEP spectrum in the range 10-100Hz. The TFN calculation assumes $F_{link} =1$, but $F_{link}$ may be as small as $1/2$ \cite{irwinhilton}. The UHF-1 v1 detectors were not measured.}
    \label{fig:nep_hist}
\end{figure}

The measured noise equivalent power should be fairly constant as a function of normal resistance fraction, and ideally, roughly flat in the range $\sim$10-100 Hz. We compare this value of the measured noise equivalent power (NEP) to an approximation of the thermal fluctuation noise (TFN), $4 k_b T_c^2 G F_{link}$ \cite{irwinhilton}. $F_{link}$ is assumed to be 1 but may be as small as $1/2$, which could explain some of the variation between the measured NEP and the TFN. See Fig. \ref{fig:noise} for an example NEP spectrum. Additional measured NEPs are shown in Fig. \ref{fig:nep_hist}.

\section{Conclusion}

Multiple versions of Simons Observatory prototype detectors have been fabricated and measured. 
Bias step and noise measurements of the NIST UHF detectors have been acquired and analyzed to inform successive generations of prototype devices. The noise measurements are consistent with expectations from bolometer thermal fluctuation noise estimates.
The second version of prototype devices from NIST meets several of the target parameters for SO, and the third version is expected to achieve sufficiently low time constants. 
Berkeley and SeeQC detectors have been measured using the IV technique with a variety of geometries. 
The most recently analyzed set of detectors contain TESes with saturation powers that are near the targets for the MF and LF bands. Future versions of these detectors will undergo similar noise and bias step measurements to those presented here. 
Final iterations of the TES designs for the MF and UHF frequencies are underway, and fabrication of the Simons Observatory detector arrays will begin in the near future.


 
\begin{acknowledgements}
This work is supported by the Simons Foundation. Laboratory Directed Research and Development (LDRD) funding from Berkeley Lab, provided by the Director, Office of Science, of the U.S. Department of Energy under Contract No. DE-AC02-05CH11231. Early Career Research Program (ECRP) program provided by the U.S. Department of Energy, Office of Science, Office of High Energy Physics, under Contract No. DE-AC02-05CH11231. Small Business Innovative Research (SBIR) program provided by the U.S. Department of Energy, Office of Science, Office of High Energy Physics, under award number DE-SC0018711 and award number HYP-DE-SC0017818. Work by NFC was supported by a NASA Space Technology Research Fellowship. MDN acknowledges support from NSF award AST-1454881.
\end{acknowledgements}


\begin{thebibliography}{99}

\bibitem{galitz}
Nicholas Galitzki, {\it et al.}, ``The Simons Observatory: instrument overview," Proc. SPIE 10708, Millimeter, Submillimeter, and Far-Infrared Detectors and Instrumentation for Astronomy IX, 1070804 (31 July 2018); https://doi.org/10.1117/12.2312985

\bibitem{orlow}
John L. Orlowski-Scherer, Ningfeng Zhu, Zhilei Xu, {\it et al.} ``Simons Observatory large aperture receiver simulation overview", Proc. SPIE 10708, Millimeter, Submillimeter, and Far-Infrared Detectors and Instrumentation for Astronomy IX, 107083X (9 July 2018); https://doi.org/10.1117/12.2312868

\bibitem{irwinhilton}
Irwin K., Hilton G. (2005) ``Transition-Edge Sensors'' In: Enss C. (eds) Cryogenic Particle Detection. Topics in Applied Physics, vol 99. Springer, Berlin, Heidelberg

\bibitem{mates}
Mates, John Arthur Benson, ``The Microwave SQUID Multiplexer'' (2011). Physics Graduate Theses and Dissertations. 9. https://scholar.colorado.edu/phys\_gradetds/9

\bibitem{henderson}
Shawn W. Henderson, {\it et al.} ``Highly-multiplexed microwave SQUID readout using the SLAC Microresonator Radio Frequency (SMuRF) electronics for future CMB and sub-millimeter surveys", Proc. SPIE 10708, Millimeter, Submillimeter, and Far-Infrared Detectors and Instrumentation for Astronomy IX, 1070819 (18 July 2018); https://doi.org/10.1117/12.2314435

\bibitem{battistelli}
Battistelli, E.S., Amiri, M., Burger, B. {\it et al. } ``Functional Description of Read-out Electronics for Time-Domain Multiplexed Bolometers for Millimeter and Sub-millimeter Astronomy'' J Low Temp Phys (2008) 151: 908. https://doi.org/10.1007/s10909-008-9772-z

\bibitem{toki}
Suzuki, A., {\it et al.} ``Commercially fabricated antenna-coupled Transition Edge Sensor bolometer detectors
for next generation Cosmic Microwave Background polarimetry experiment'' J Low Temp Phys This Special Issue (2019)


\bibitem{li}
Li, D., Austermann, J. E., Beall, J. A. {\it et al.} ``AlMn Transition Edge Sensors for Advanced ACTPol'', J Low Temp Phys (2016) 184:66.
https://doi.org/10.1007/s10909-016-1526-8

\bibitem{choi}
Choi, S.K., Austermann, J., Beall, J.A. {\it et al.} ``Characterization of the Mid-Frequency Arrays for Advanced ACTPol'' J Low Temp Phys (2018) 193: 267. https://doi.org/10.1007/s10909-018-1982-4

\bibitem{koopman}
Koopman, B., Cothard, N. F., Choi, S. K., {\it et al.} ``Advanced ACTPol Low Frequency Array:
Readout and Characterization of Prototype 27 and 39 GHz Transition Edge Sensors'' Journal of
Low Temperature Physics (May 11, 2018); doi:10.1007/s10909-018-1957-5, arXiv:1711.02594

\bibitem{two-fluid}
Irwin, K. D., Hilton G. C., Wollman D. A., Martinis J. M. ``Thermal-response time of superconducting transition-edge microcalorimeters'' J Appl Phys (1998) 83:3978. https://doi.org/10.1063/1.367153

\bibitem{C_PdAu}
Laufer, P. M., Papaconstantopoulos, D. A., ``Tight-binding coherent-potential-approximation study of the electronic states of palladium–noble-metal alloys'' Phys Rev B (1987) 35:9019. https://doi.org/10.1103/PhysRevB.35.9019

\bibitem{so_forecast}
Ade, P., {\it et al} ``The Simons Observatory: science goals and forecasts'' Journal of Cosmology and Astroparticle Physics (2019) doi:10.1088/1475-7516/2019/02/056

\end{thebibliography}
\end{document}